\documentclass[11pt,letterpaper]{article}
\usepackage{systeme} 
\usepackage[utf8]{inputenc}
\usepackage{multirow} 
\usepackage{booktabs} 
\usepackage[english]{babel}
\usepackage{amsmath}
\usepackage{amsfonts}
\usepackage{amssymb}
\usepackage{graphicx}
\usepackage[small,bf]{caption} 
\usepackage[left=3cm,right=3cm,top=2cm,bottom=3cm]{geometry}
\usepackage{framed} 
\usepackage{color} 
\usepackage{wrapfig}\definecolor{shadecolor}{RGB}{220,220,220} 
\usepackage{float} 
\usepackage{array} 
\usepackage{caption} 
\usepackage{wasysym} 
\author{Jorge Pinochet}
\title{\textbf{Calculating a journey to Mars with introductory physics}}
\begin{document}

\author{Jorge Pinochet$^{*}$\\ \\
 \small{$^{*}$\textit{Facultad de Ciencias Básicas, Departamento de Física. }}\\
  \small{\textit{Centro de Desarrollo de Investigación CEDI-UMCE,}}\\
 \small{\textit{Universidad Metropolitana de Ciencias de la Educación,}}\\
 \small{\textit{Av. José Pedro Alessandri 774, Ñuñoa, Santiago, Chile.}}\\
 \small{e-mail: jorge.pinochet@umce.cl}\\}

\date{}
\maketitle

\begin{center}\rule{0.9\textwidth}{0.1mm} \end{center}
\begin{abstract}
\noindent The problem of a crewed mission to Mars provides a compelling and realistic context for teaching introductory physics. In this work, we present a coherent set of calculations describing a round-trip journey between Earth and Mars, using only mathematical tools accessible at the high-school level. The treatment integrates key topics that are typically taught separately —such as Newtonian gravitation, energy, circular motion, and basic astrodynamics— into a single coherent framework. The analysis includes Hohmann transfer orbits, travel times, launch conditions, and mission constraints, illustrating how a complex real-world problem can be transformed into an effective classroom resource. This work aims to support teachers and students in exploring interplanetary travel as an accessible and motivating application of fundamental physics.\\ \\

\noindent \textbf{Keywords}: Universal gravitation, astrodynamics, interplanetary travel.

\begin{center}\rule{0.9\textwidth}{0.1mm} \end{center}
\end{abstract}

\maketitle

\section{Introduction}

For at least 70 years, scientists and engineers have dreamed of sending a crewed mission to Mars. Although the technical, logistical, financial, and human challenges are formidable, agencies such as NASA (National Aeronautics and Space Administration) and ESA (European Space Agency), together with private companies such as SpaceX, Blue Origin, and Virgin Galactic, have actively pursued this goal in recent decades.\\

In this context, travel to Mars has become a topic of great interest, not only for the scientific community but also for the general public. From the standpoint of physics, one of the main challenges is to perform the calculations required to send a crewed spacecraft to Mars and bring it back. While there exist popular science resources with simplified treatments, these often omit aspects relevant for a more complete understanding. On the other hand, the technical literature presents a level of mathematical complexity that makes it less accessible to students and teachers in training [1–3].\\

From an educational perspective, the problem of a mission to Mars can be used as an integrative activity that connects topics usually treated separately in introductory courses. It provides a realistic context in which students apply concepts, theories, and laws, fostering a deeper conceptual understanding of physics and astrodynamics. Moreover, the use of accessible mathematical tools allows its implementation both in secondary education and in early undergraduate levels.\\

The objective of this work is to develop the essential calculations for a crewed mission to Mars in a context that allows the articulation of fundamental concepts of introductory physics. In this way, the problem is presented not only as a technical exercise but also as an educational resource that promotes the understanding of the physics of interplanetary travel. The adopted approach, based on mathematical tools that do not exceed high-school algebra, facilitates its implementation in the classroom.

\section{Theoretical framework}

Successfully completing a space mission in which a group of astronauts visits Mars is an ambitious objective that requires enormous technical and financial efforts. Among the technical aspects, one of the most important is performing the necessary calculations to send a crewed spacecraft and bring it back. These calculations require the contribution of various areas of physics, such as Newtonian mechanics, gravitation, rotational kinematics, and astrodynamics. In this section, we present a concise summary of all the tools required to carry out a crewed mission to Mars.

\subsection{Law of gravitation}

As illustrated in Fig.~1, this law establishes that a spherical celestial body of mass $M$ and a particle of mass $m$ exert an attractive force on each other that follows the equation [4]

\begin{equation}
F = \frac{G M m}{r^2},
\end{equation}

where $F$ is the magnitude of the attractive force, $r$ is the distance between the particle and the center of the celestial body, and $G$ is the gravitational constant. Since $F$ is a conservative force, it has an associated potential energy $U$ given by [4]

\begin{equation}
U = -\frac{G M m}{r}.
\end{equation}

Therefore, the total mechanical energy $E$ of a particle subjected to gravity is the sum of its kinetic energy $K$ and its potential energy [5]

\begin{equation}
E = K + U = \frac{1}{2} m v^2 - \frac{G M m}{r},
\end{equation}

where $v$ is the speed of the particle. Since the particle is subjected to a centripetal force of magnitude $m v^2 / r$, provided by gravity, we have $\frac{m v^2}{r} = \frac{G M m}{r^2}$. Multiplying both sides by $1/2$, we obtain

\begin{equation}
\frac{1}{2} m v^2 = \frac{1}{2} \frac{G M m}{r}.
\end{equation}

The term on the left-hand side is $K$, and the term on the right-hand side is $-U/2$, so that $K = -U/2$. Substituting this result into Eq. (3), we obtain

\begin{equation}
E = -\frac{U}{2} + U = \frac{U}{2} = -\frac{G M m}{2 r} < 0.
\end{equation}

\begin{figure}[H]
  \centering
    \includegraphics[width=0.3\textwidth]{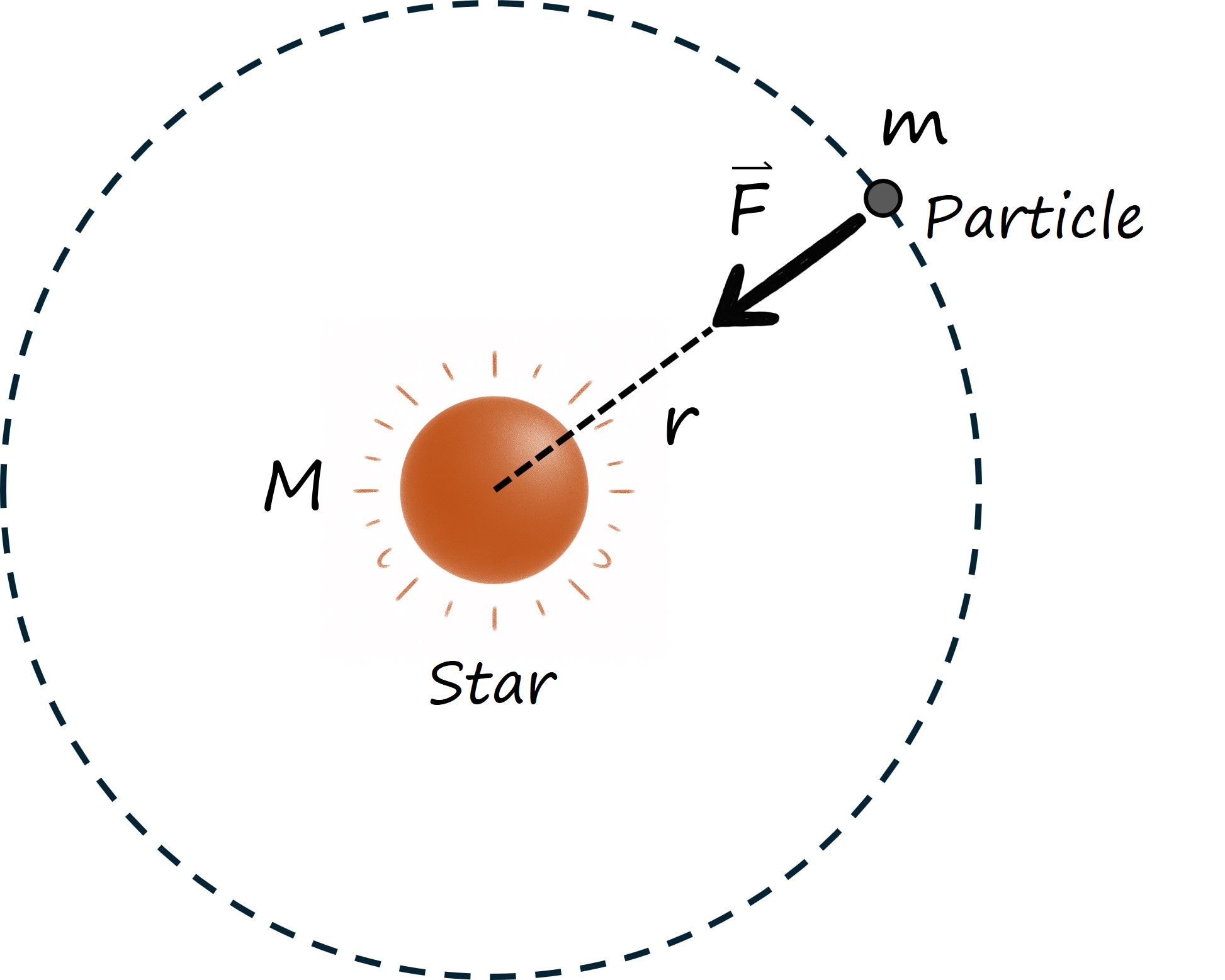}
  \caption{Particle of mass $m$ describing a circular orbit around a massive body of mass $M \gg m$.}
\end{figure}

Thus, the particle is in a bound state, since its total energy in a circular orbit is negative. According to Kepler's third law, this is a particular case of an elliptical orbit, which is the most general situation. We can extend Eq. (5) to elliptical orbits by replacing the radius $r$ with the semi-major axis $a$ of the ellipse, yielding [6]

\begin{equation}
E = -\frac{G M m}{2 a}.
\end{equation}

We see that, for an elliptical orbit, the particle is also in a bound state. Figure~2 shows an ellipse along with its defining parameters: the semi-major axis $a$, the semi-minor axis $b$, and the foci $F_1$ and $F_2$. Kepler's first law also states that when a particle follows an elliptical orbit around a celestial body, the latter is located at one of the foci.\\

\begin{figure}[H]
  \centering
    \includegraphics[width=0.35\textwidth]{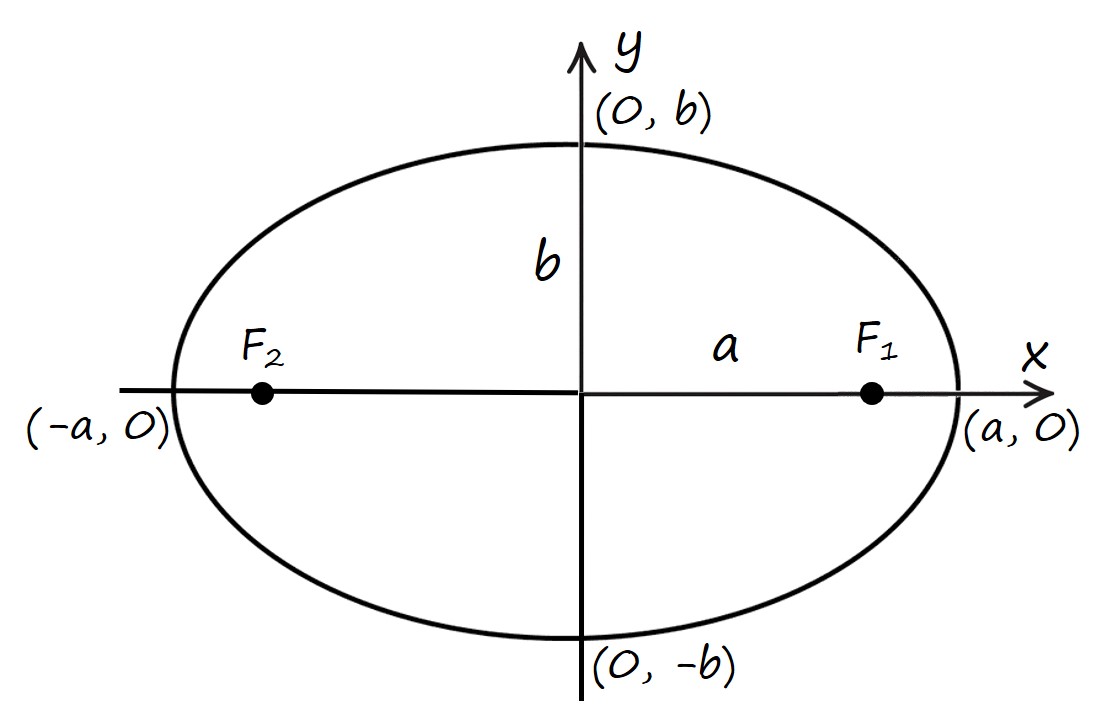}
  \caption{Main parameters that characterize an ellipse.}
\end{figure}

Finally, we must consider Kepler's third law. For the case of a circular orbit, this law can be derived from Eq. (4) by expressing the speed as $v = \frac{2\pi r}{T}$, where $T$ is the orbital period, so that $\left(\frac{2\pi r}{T}\right)^2 = \frac{G M}{r}$. After some algebra, we obtain

\begin{equation}
T^2 = \frac{4\pi^2}{G M} r^3.
\end{equation}

We can generalize this law by replacing the radius $r$ with the semi-major axis $a$ of the ellipse:

\begin{equation}
T = 2\pi \sqrt{\frac{a^3}{G M}}.
\end{equation}

\subsection{Hohmann transfer orbit}

An important application of the ideas developed in the previous section is the so-called \textit{Hohmann transfer orbit} (HTO), which is a space maneuver that transfers a spacecraft from one circular orbit to another using the minimum amount of energy [3]. This is crucial, since fuel availability is one of the most critical aspects of spaceflight.\\

\begin{figure}[H]
  \centering
    \includegraphics[width=0.6\textwidth]{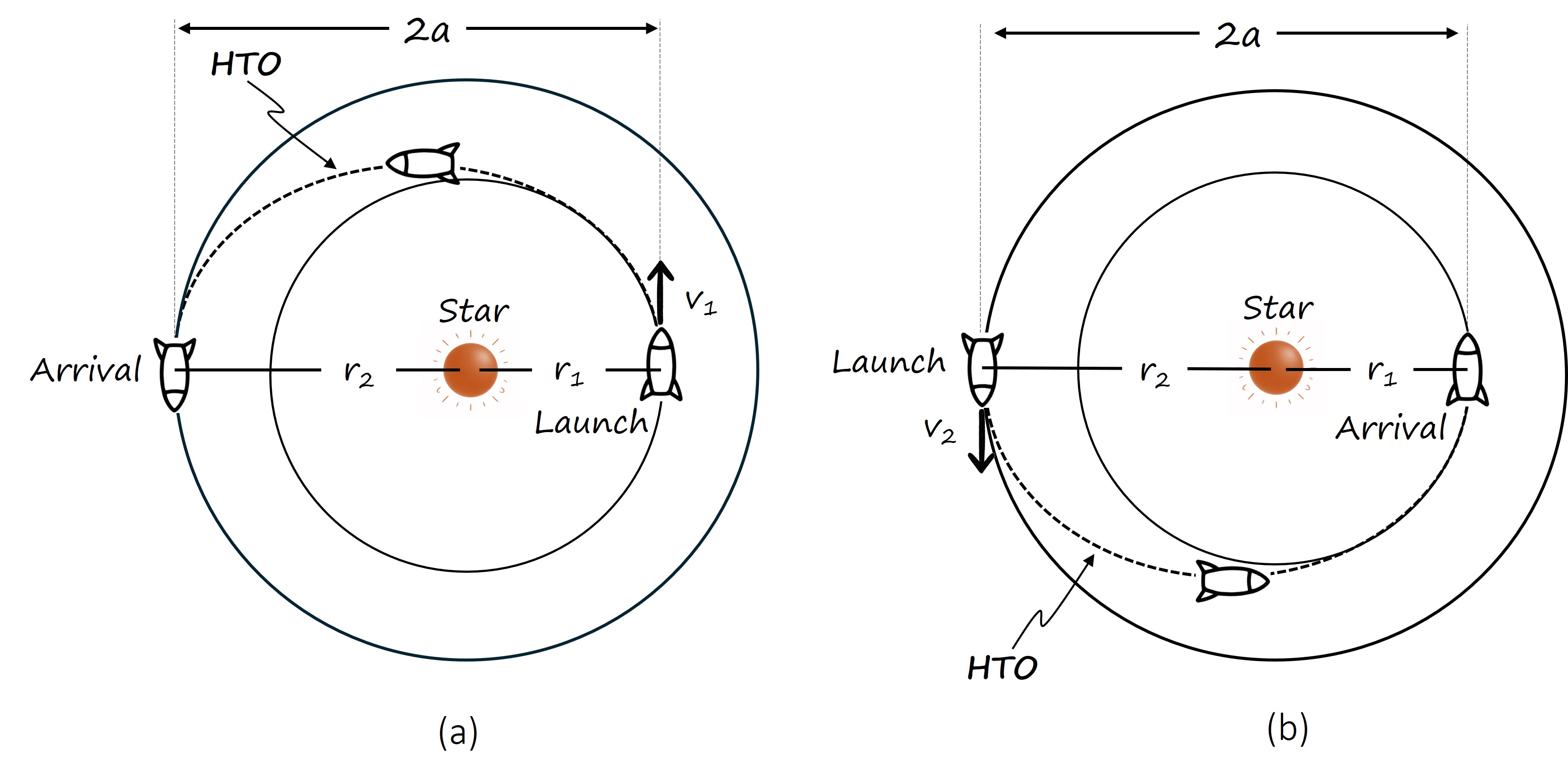}
  \caption{Parameters involved in an HTO. (a) A spacecraft moves from an orbit $r_1$ to another with $r_2 > r_1$. (b) A spacecraft moves from an orbit $r_2$ to another with $r_1 < r_2$.}
\end{figure}

A fundamental assumption of this maneuver is that the spacecraft can be considered a particle, since its mass is negligible compared to that of the central celestial body. The HTO corresponds to half of an elliptical orbit that is tangent to both the initial orbit to be left and the final orbit to be reached [3]. Figure~3 shows the main aspects involved in an HTO, where a spacecraft of mass $m$ orbiting a star of mass $M\gg m$ moves from an orbit of radius $r_1$ to one of radius $r_2 > r_1$. At $r_1$, the spacecraft fires its engines, receiving an impulse with velocity $v_1$, which provides the additional energy required to reach orbit $r_2$ following an HTO.\\

\begin{figure}[H]
  \centering
    \includegraphics[width=0.25\textwidth]{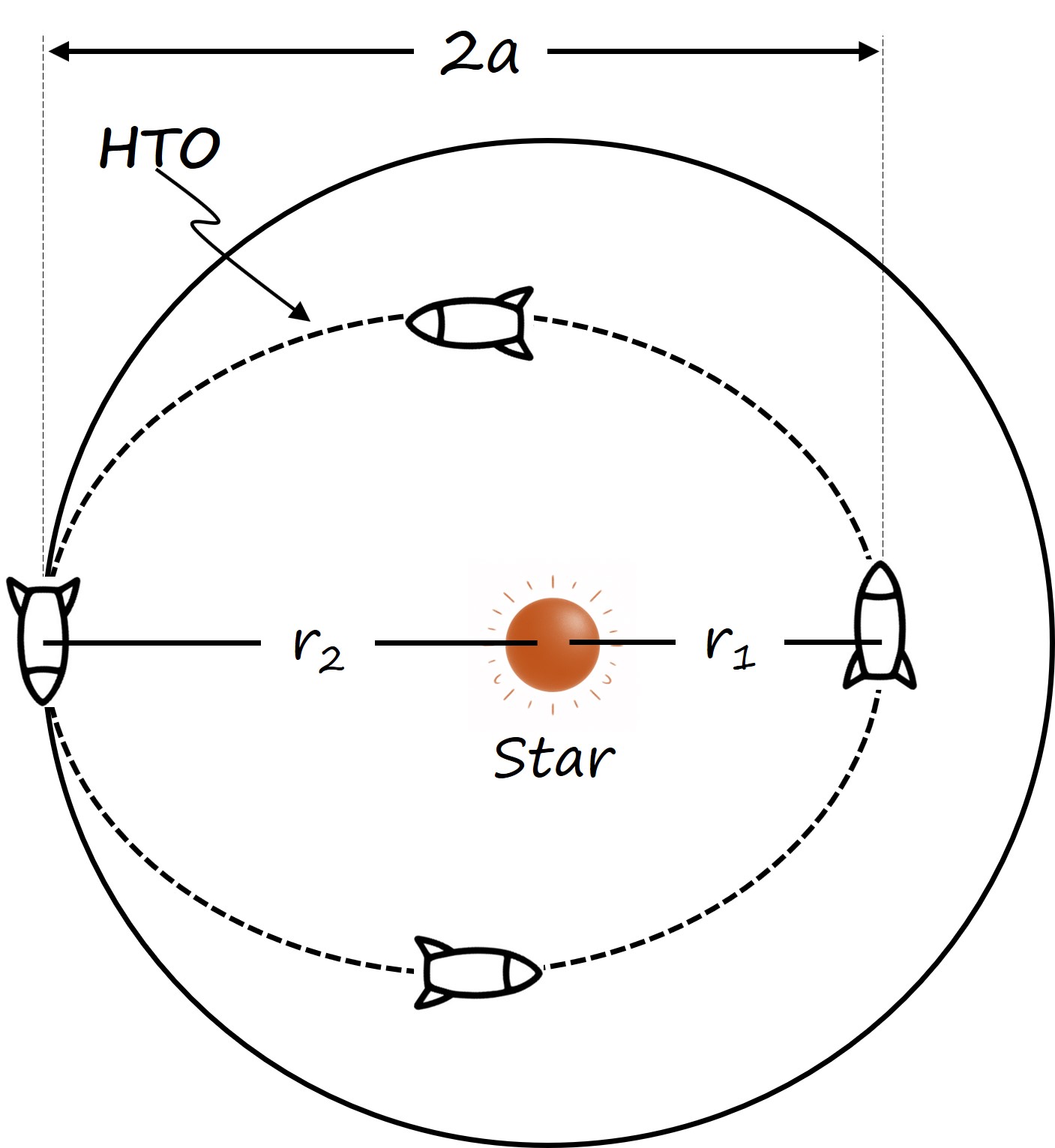}
  \caption{HTO that the spacecraft would follow if it did not land on Mars.}
\end{figure} 
According to Fig.~3(a), the velocity $v_1$ required for the spacecraft to move from orbit $r_1$ to $r_2$ can be calculated using

\begin{equation}
-\frac{G M m}{2 a} = \frac{1}{2} m v_1^2 - \frac{G M m}{r_1}.
\end{equation}

Here, $a$ is the semi-major axis of the HTO, related to the initial and final orbital radii by $r_1 + r_2 = 2a$. Substituting this relation into the previous equation and solving for the speed:

\begin{equation}
v_1 = \sqrt{2 G M \left( \frac{1}{r_1} - \frac{1}{r_1 + r_2} \right)}.
\end{equation}

An HTO also allows the spacecraft to move from $r_2$ to $r_1$. In this case, the spacecraft must fire its engines in the direction opposite to its motion in order to lose energy and fall into a lower orbit. The equation for $v_2$ is:

\begin{equation}
v_2 = \sqrt{2 G M \left( \frac{1}{r_2} - \frac{1}{r_1 + r_2} \right)}.
\end{equation}

If the spacecraft were simply to describe a free HTO, without encountering any planet along its trajectory, it would follow the elliptical orbit shown in Fig. 4.

\subsection{Sphere of influence}

A \textit{sphere of influence} (SOI) is a spheroidal region around a celestial body where its gravitational influence dominates over that of another celestial body. Therefore, the SOI has no physical existence and is only a mathematical concept useful for performing calculations in astrodynamics [1]. Given a body of mass $m$ and another body of mass $M > m$, the general equation is $R_{\text{SOI}} = a \left( \frac{m}{M} \right)^{2/5}$ [2], where $a$ is the semi-major axis of the orbit. If the orbit is circular, then $a = r$ and

\begin{equation}
R_{\text{SOI}} = r \left( \frac{m}{M} \right)^{2/5}.
\end{equation}

\subsection{Rotational kinematics}

Let us consider a particle undergoing uniform circular motion along a trajectory of radius $r$. If the particle moves counterclockwise from an initial angle $\theta_0$ at time $t_0$ to a final angle $\theta$ at time $t$, the angular velocity is $ \omega = \frac{\theta - \theta_0}{t - t_0}$ [7] (see Fig. 5). Taking $t_0 = 0$, we obtain

\begin{equation}
\omega = \frac{\theta - \theta_0}{t}.
\end{equation}

\begin{figure}[H]
  \centering
    \includegraphics[width=0.3\textwidth]{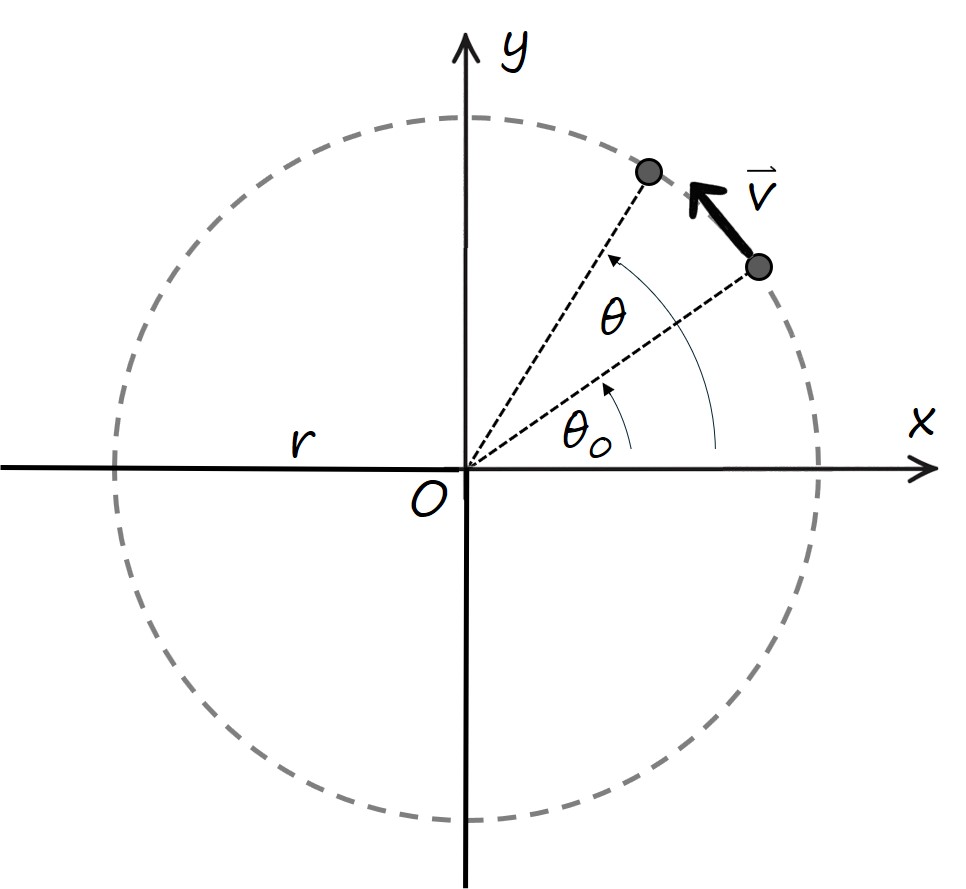}
  \caption{Parameters involved in uniform circular motion.}
\end{figure}

Since the motion is uniform, $\omega$ is constant, and thus

\begin{equation}
\theta = \theta_0 + \omega t.
\end{equation}

In the calculations of the mission to Mars, we assume that Earth and Mars follow circular orbits with constant angular velocity. Therefore, these equations contain all the necessary information on rotational kinematics.

\section{Calculating the trip to Mars}

The ideas developed in the previous section constitute our toolkit for performing the calculations of the mission to Mars. The first part of the mission, corresponding to the outbound journey, requires launching the spacecraft from Earth in such a way that it is inserted into Mars' trajectory at the correct moment so that the red planet is there. The second part, corresponding to the return journey, requires launching the spacecraft such that it is inserted into Earth's trajectory at the correct moment so that our planet is there. This is similar to firing a bullet from a moving rifle at a moving target, making it a complex problem.\\

When the spacecraft completes the outbound journey, it must enter orbit around Mars, and from there, the astronauts will descend to the planet. When the spacecraft completes the return journey, it must enter orbit around Earth, and from there, the astronauts will descend to our planet. Table 1 summarizes the ten parameters involved in the calculations, along with their symbols and values in SI units.\\

\begin{table}[h]
\centering
\footnotesize
\begin{tabular}{lll}
\hline
\textbf{Parameter} & \textbf{Symbol} & \textbf{Value (SI)} \\
\hline
Mass of the Sun & $M_{\odot}$ & $1.99 \times 10^{30}\,\mathrm{kg}$ \\
Mass of the Earth & $M_{\oplus}$ & $5.98 \times 10^{24}\,\mathrm{kg}$ \\
Mass of Mars & $M_{\mars}$ & $6.56 \times 10^{23}\,\mathrm{kg}$ \\
Radius of Earth's orbit & $r_{\oplus}$ & $1.49 \times 10^{11}\,\mathrm{m}$ \\
Radius of Mars' orbit & $r_{\mars}$ & $2.28 \times 10^{11}\,\mathrm{m}$ \\
Orbital period of Earth & $T_{\oplus}$ & $3.2 \times 10^{7}\,\mathrm{s}$ \\
Orbital period of Mars & $T_{\mars}$ & $5.9 \times 10^{7}\,\mathrm{s}$ \\
Orbital speed of Earth & $v_{\oplus}$ & $2.98 \times 10^{4}\,\mathrm{m\,s^{-1}}$ \\
Orbital speed of Mars & $v_{\mars}$ & $2.40 \times 10^{4}\,\mathrm{m\,s^{-1}}$ \\
Gravitational constant & $G$ & $6.67 \times 10^{-11}\,\mathrm{N\,m^{2}\,kg^{-2}}$ \\
\hline
\end{tabular}
\caption{Parameters involved in the calculations of the mission to Mars}
\end{table}

For simplicity, in the calculations that follow we will assume that the spacecraft departs from an orbit around Earth and arrives into an orbit around Mars, omitting the details related to the maneuvers required to remain in such orbits and to ascend or descend from the surfaces of Mars and Earth. We will also assume that the orbits of Earth and Mars are circular, with the Sun at the center.\\

In this scenario, we begin by analyzing the transfer orbit that will allow the astronauts to travel from Earth to Mars, and we conclude with the transfer orbit that will bring them back.

\subsection{Spheres of influence of Earth and Mars}

To perform the calculations that allow the successful completion of the mission to Mars, we must show that, during its journey, the spacecraft behaves as if it were only under the gravitational influence of the Sun. That is, we must demonstrate that the spheres of influence of Earth and Mars are negligible compared to the gravitational influence of the Sun. To this end, it suffices to determine the ratio $R_{\mathrm{SOI}}/r$, which provides the radius of influence of Earth and Mars relative to their distances from the Sun. For our planet, introducing into Eq.~(12) the Earth's mass given in Table 1, we obtain

\begin{equation}
\frac{R_{\mathrm{SOI}}}{r_{\oplus}} = \left( \frac{M_{\oplus}}{M_{\odot}} \right)^{2/5} \simeq 6.2 \times 10^{-3}.
\end{equation}

This result indicates that Earth's sphere of influence is only $0.62\%$ of its distance to the Sun. Similarly, according to Eq.~(12) and Table 1, for Mars we have

\begin{equation}
\frac{R_{\mathrm{SOI}}}{r_{\mars}} = \left( \frac{M_{\mars}}{M_{\odot}} \right)^{2/5} \simeq 2.53 \times 10^{-3}.
\end{equation}

This result indicates that Mars' sphere of influence is $0.25\%$ of its distance to the Sun. Therefore, as expected, the spacecraft will follow a heliocentric trajectory determined almost exclusively by the Sun's gravitational influence, which greatly simplifies the calculations.

\subsection{The outbound journey}

\textbf{Initial velocity:} Fig. 6 illustrates the journey from Earth to Mars. As shown, Mars must be located ahead of Earth by an angle $\theta_{1}$ at the moment of launch. This condition follows from the finite transfer time and the fact that Mars, moving more slowly than Earth, advances along its orbit during this time.

\begin{figure}[H]
  \centering
    \includegraphics[width=0.5\textwidth]{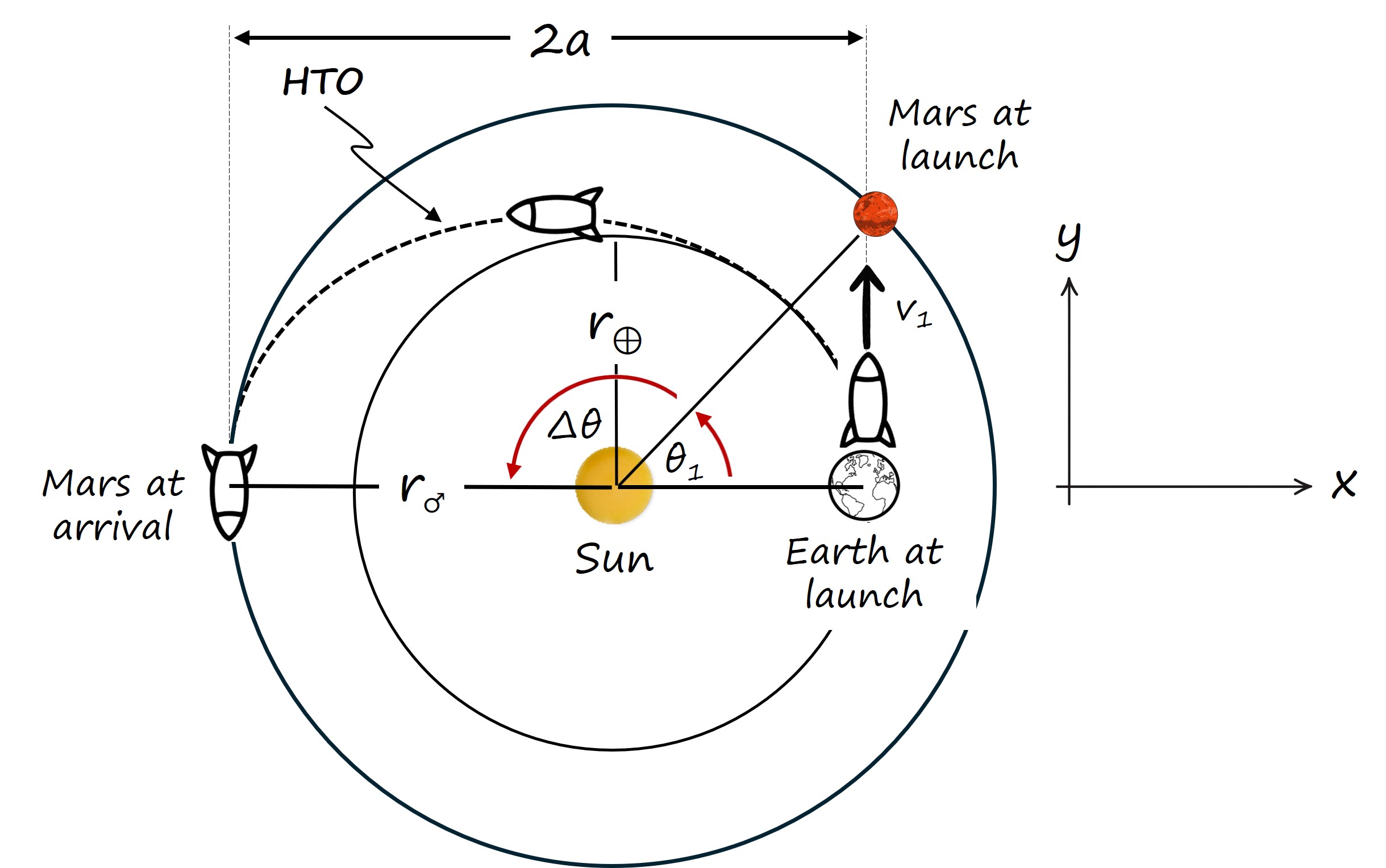}
  \caption{HTO on the spacecraft's journey from Earth to Mars.}
\end{figure}

Fig. 6 is analogous to Fig.~3(a), except that now $r_1 = r_{\oplus}$ and $r_2 = r_{\mars}$, where $v_1$ is the speed of the spacecraft departing from Earth. Additionally, the figure shows that the major axis of the ellipse is given by $2a = r_{\oplus} + r_{\mars}$. Introducing the corresponding parameters from Table 1 into Eq.~(10), we obtain the speed of the spacecraft when it fires its engines from our planet:

\begin{equation}
v_1 = \sqrt{2GM_{\odot} \left( \frac{1}{r_{\oplus}} - \frac{1}{r_{\oplus} + r_{\mars}} \right)} \simeq 3.28 \times 10^{4}\,\mathrm{m\,s^{-1}}.
\end{equation}

Since, at the moment of firing its engines, the spacecraft is in orbit around Earth, it moves together with it and shares its orbital velocity $v_{\oplus}$. This means that $v_1 = v_{\oplus} + \Delta v_1$, where $\Delta v_1$ is the velocity increment required for the spacecraft to begin the journey to Mars. Thus,

\begin{equation}
\Delta v_1 = v_1 - v_{\oplus} \simeq 3.0 \times 10^{3}\,\mathrm{m\,s^{-1}}.
\end{equation}

\textbf{Travel time:} To determine the time it takes for the spacecraft to travel the distance between Earth and Mars, we use Kepler's third law, Eq.~(8). Since a transfer orbit is half an elliptical orbit, the time $\Delta t_1$ required to move from one orbit to another is half the period, i.e., $t_1 = T/2$. Taking $a = (r_{\oplus} + r_{\mars})/2$ and introducing the values from Table 1, we obtain

\begin{equation}
t_1 = \pi \sqrt{\frac{(r_{\oplus} + r_{\mars})^{3}}{8GM_{\odot}}} \simeq 2.2 \times 10^{7}\,\mathrm{s} \simeq 255\,\mathrm{days}.
\end{equation}

\textbf{Initial angle:} As shown in Fig.~6, at the moment of launch, Mars' orbital radius forms an angle $\theta_1$ with respect to Earth's orbital radius. According to Eq.~(13), to find $\theta_1$ we use the angular velocity of Mars in its orbital motion, given by $\omega_{\mars} = 360^{\circ}/T_{\mars}$ (see Table 1). During the time the spacecraft travels from Earth to the meeting point with Mars ($t_1 \simeq 255$ days), Mars sweeps an angle $\Delta \theta = \omega_{\mars} t_1 \simeq 134^{\circ}$. Therefore, the required angle is

\begin{equation}
\theta_1 = 180^{\circ} - \Delta \theta \simeq 46^{\circ}.
\end{equation}

\subsection{The return journey}

\textbf{Initial velocity:} Figure 7 illustrates the journey from Mars to Earth. Once again, it is observed that Mars must be located ahead of Earth by an angle $\theta_{1}$ at the moment of launch. Fig. 7 is analogous to Fig.~3(b), except that now $r_1 = r_{\oplus}$ and $r_2 = r_{\mars}$, where $v_2$ is the speed of the spacecraft departing from Mars. Recalling that $2a = r_{\oplus} + r_{\mars}$ and introducing the values from Table 1 into Eq.~(11), we obtain

\begin{equation}
v_2 = \sqrt{2GM_{\odot} \left( \frac{1}{r_{\mars}} - \frac{1}{r_{\oplus} + r_{\mars}} \right)} \simeq 2.15 \times 10^{4}\,\mathrm{m\,s^{-1}}.
\end{equation}

\begin{figure}[H]
  \centering
    \includegraphics[width=0.5\textwidth]{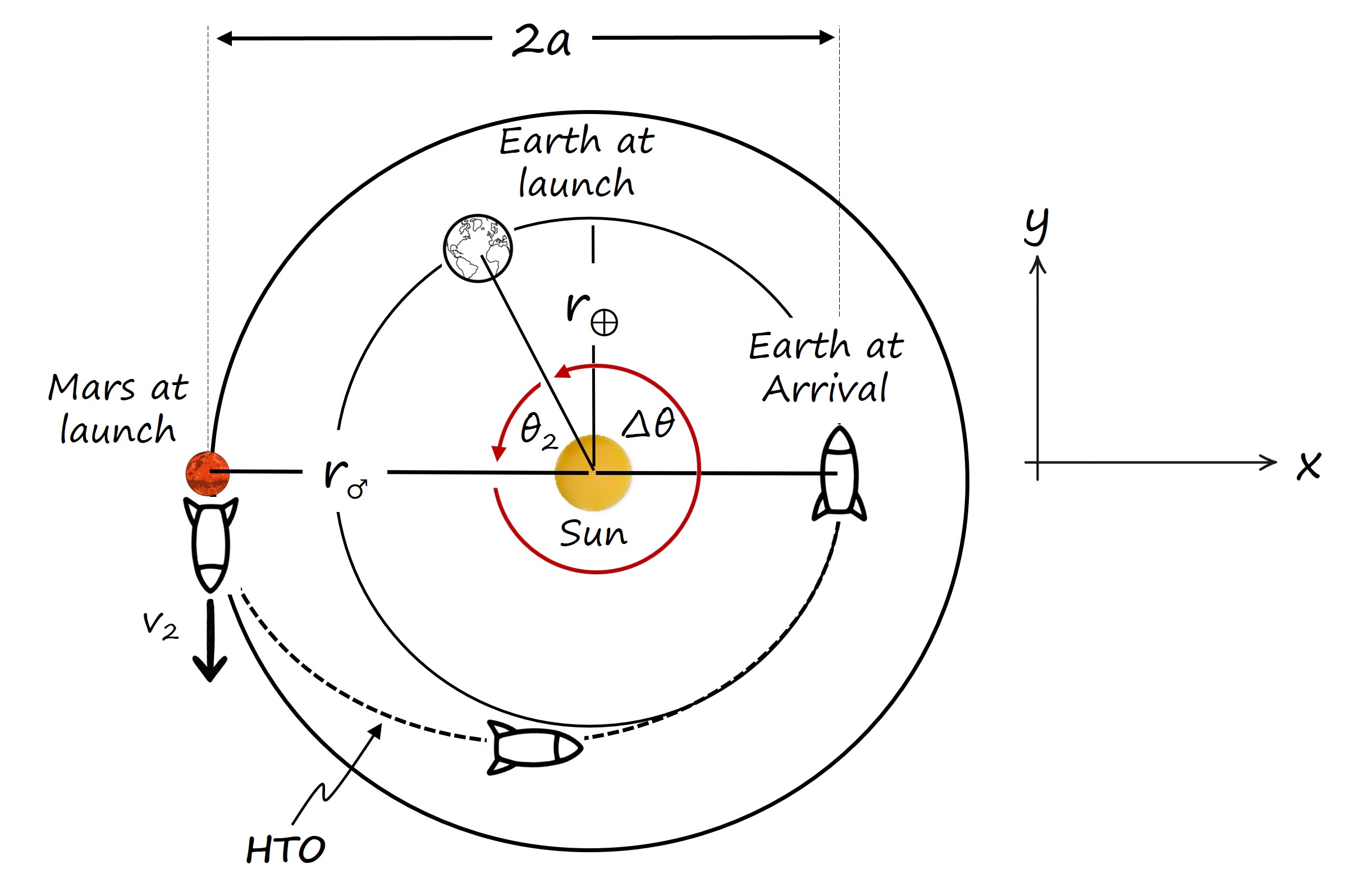}
  \caption{HTO on the spacecraft's journey from Mars to Earth.}
\end{figure}

Since the spacecraft is in orbit around Mars, it moves together with the planet and shares its orbital velocity $v_{\mars}$. This implies that the spacecraft must fire its engines in the opposite direction to $v_{\mars}$. As a result, $v_2$ is the sum of $v_{\mars}$ and the negative velocity change $-\Delta v_2$ required to initiate the return journey, i.e., $v_2 = v_{\mars} - \Delta v_2$, so that

\begin{equation}
\Delta v_2 = v_{\mars} - v_2 \simeq 2.5 \times 10^{3}\,\mathrm{m\,s^{-1}}.
\end{equation}

\textbf{Travel time:} Since the trajectory, and therefore the distance traveled during the return journey, is the same as in the outbound journey, the travel times must also be the same:

\begin{equation}
t_2 = t_1 \simeq 255\,\mathrm{days}.
\end{equation}

\textbf{Initial angle:} As shown in Fig.~7, at the moment of launching the spacecraft back to Earth, the orbital radii of Mars and Earth form an angle $\theta_2$, where Earth is located behind Mars. To calculate this angle, we need Earth's angular velocity in its orbital motion, given by $\omega_{\oplus} = 360^{\circ}/T_{\oplus}$. During the time the spacecraft travels to the meeting point with Earth (255 days), Earth sweeps an angle $\Delta \theta = \omega_{\oplus} t_2 \simeq 248^{\circ}$. According to Fig.~7, $\Delta \theta = 180^{\circ} + \theta_2$, hence

\begin{equation}
\theta_2 = \Delta \theta - 180^{\circ} \simeq 68^{\circ}.
\end{equation}

\textbf{Waiting time on Mars:} As shown in Fig.~7, Earth appears to be behind Mars. However, since Earth's orbital period is shorter than that of Mars, Earth moves faster. Thus, although Earth appears to lag behind Mars by $68^{\circ}$, it is actually ahead by $360^{\circ} - 68^{\circ} = 292^{\circ}$. This implies that, to initiate the return journey, the astronauts must wait a time interval $\Delta t$ until Earth is ahead of Mars by $292^{\circ}$.\\

To calculate $\Delta t$, we write the kinematic equations for the angular positions of Earth and Mars starting from the moment the spacecraft is launched from Earth: $\theta_{\oplus} = \omega_{\oplus} \Delta t$, $\theta_{\mars} = \omega_{\mars} \Delta t + \theta_1$. The condition for Earth to be ahead of Mars by $292^{\circ}$ is $\theta_{\oplus} - \theta_{\mars} = 292^{\circ}$. Imposing these conditions, from Eq.~(14) we obtain

\begin{equation}
\theta_{\oplus} - \theta_{\mars} = 292^{\circ} = \omega_{\oplus} \Delta t - (\omega_{\mars} \Delta t + \theta_1).
\end{equation}

Solving for the time,

\begin{equation}
\Delta t \simeq \frac{292^{\circ} + \theta_1}{\omega_{\oplus} - \omega_{\mars}} \simeq 6.6 \times 10^{7}\,\mathrm{s} \simeq 764\,\mathrm{days}.
\end{equation}

However, this interval includes the time $t_2 = t_1 \simeq 255$ days corresponding to the transfer orbit from Earth to Mars. Therefore, the waiting time $t_e$ is

\begin{equation}
t_e = \Delta t - t_1 = 764\,\mathrm{days} - 255\,\mathrm{days} \simeq 509\,\mathrm{days}.
\end{equation}

In summary, the complete mission lasts a total time of $255\,\mathrm{days} + 509\,\mathrm{days} + 255\,\mathrm{days} \simeq 1019\,\mathrm{days}$, which, approximately, corresponds to $2.8$ years.

\section{Comparison with other approaches}

Despite the strong public interest that space travel generates, and its scientific and technological importance, this topic has received limited attention in physics education journals oriented toward practical teaching, such as Physics Education and The Physics Teacher, which are mainly focused on secondary education and the early years of undergraduate study. In particular, relatively few contributions present analytically accessible treatments suitable for these levels. A relevant contribution in this direction is the work by Stinner and Begoray, \textit{Journey to Mars: the physics of travelling to the red planet} [8], which shares some similarities with the approach developed in this paper.\\

To position the contribution of the present study, Table 2 provides a brief comparison between both approaches. In both cases, the journey to Mars is used as an integrative context for fundamental concepts such as gravitation and orbital motion; however, while Stinner and Begoray rely on an interactive computer program to explore different mission scenarios, the present work develops a coherent sequence of accessible analytical calculations, leading to explicit quantitative results (velocities, travel times, and launch conditions) using only mathematical tools at the high-school level.\\

\begin{table}[h]
\centering
\footnotesize
\begin{tabular}{lll}
\hline
\textbf{Stinner and Begoray (2005)} & & \textbf{This work} \\
\hline
Interactive computer program (ICP) & & Analytical step-by-step development \\
Qualitative / semi-quantitative approach & & High-school level algebra \\
Exploration of multiple scenarios & & Structured sequence of calculations \\
Conceptual understanding via simulation & & Explicit quantitative results ($\Delta v$, $t$, $\theta$) \\
Requires computational resources & & No technology required \\
\hline
\multicolumn{3}{l}{Common feature: Mars mission as an integrative physics context} \\
\multicolumn{3}{l}{Main contribution: analytical and accessible alternative for classroom implementation} \\
\hline
\end{tabular}
\caption{Comparison between Stinner and Begoray (2005) and the present work.}
\end{table}

In this sense, the main contribution of this work is to provide a pedagogical alternative centred on analytical reasoning and conceptual integration, complementing previous simulation-based approaches and offering a resource that can be readily implemented in classroom settings, particularly where computational tools are not available. Within this framework, particular emphasis is placed on the pedagogical value of integrating traditionally separated topics into a single coherent structure.

\section{Final remarks}

Table 3 summarizes the main results obtained for the outbound and return mission to Mars, including the velocity increments, travel times, and launch angles. Beyond their numerical values, these results provide an opportunity to discuss in the classroom various fundamental aspects of physics and astrodynamics.\\

\begin{table}[h]
\centering
\footnotesize
\begin{tabular}{lll}
\hline
\textbf{Outbound Journey} & & \textbf{Return Journey} \\
\hline
$\Delta v_1 = 3.0 \times 10^{3}\,\mathrm{m\,s^{-1}}$ & & $\Delta v_2 = 2.5 \times 10^{3}\,\mathrm{m\,s^{-1}}$ \\
$t_1 = 255\,\mathrm{days}$ & & $t_2 = 255\,\mathrm{days}$ \\
$\theta_1 = 46^{\circ}$ & & $\theta_2 = 68^{\circ}$ \\
\hline
\multicolumn{3}{l}{Waiting time = $509$ days} \\
\multicolumn{3}{l}{Total mission duration = $2.8$ years} \\
\hline
\end{tabular}
\caption{Parameters involved in the mission to Mars.}
\end{table}

In particular, the fact that travel times are on the order of hundreds of days helps to appreciate the scales involved in interplanetary travel, while the values of $\Delta v$ highlight the energetic demands of such missions. Likewise, the need to wait for an extended period on Mars before initiating the return journey clearly illustrates the importance of orbital synchronization and the dynamical nature of the Earth--Mars system.\\

From a pedagogical perspective, these results can be used to promote the physical interpretation of quantities, the analysis of orders of magnitude, and the discussion of the assumptions adopted, such as the approximation of circular orbits or the consideration of heliocentric trajectories. In this way, the problem of the mission to Mars not only allows for the application of theoretical concepts, but also fosters a more integrated understanding of the physics involved in real situations.

\section{Appendix: Propulsion of space rockets}

\renewcommand{\thefigure}{A\arabic{figure}}
\setcounter{figure}{0}

Let us consider the situation illustrated in Fig.~A1, where a space rocket moves in the radial direction against Earth's gravity. The rocket has initial mass $M + dm$ and velocity $V$ relative to Earth. Under these conditions, the initial momentum of the rocket is

\begin{equation}
p_i = (M + dm)V.
\tag{A1}
\end{equation}

\begin{figure}[H]
  \centering
    \includegraphics[width=0.6\textwidth]{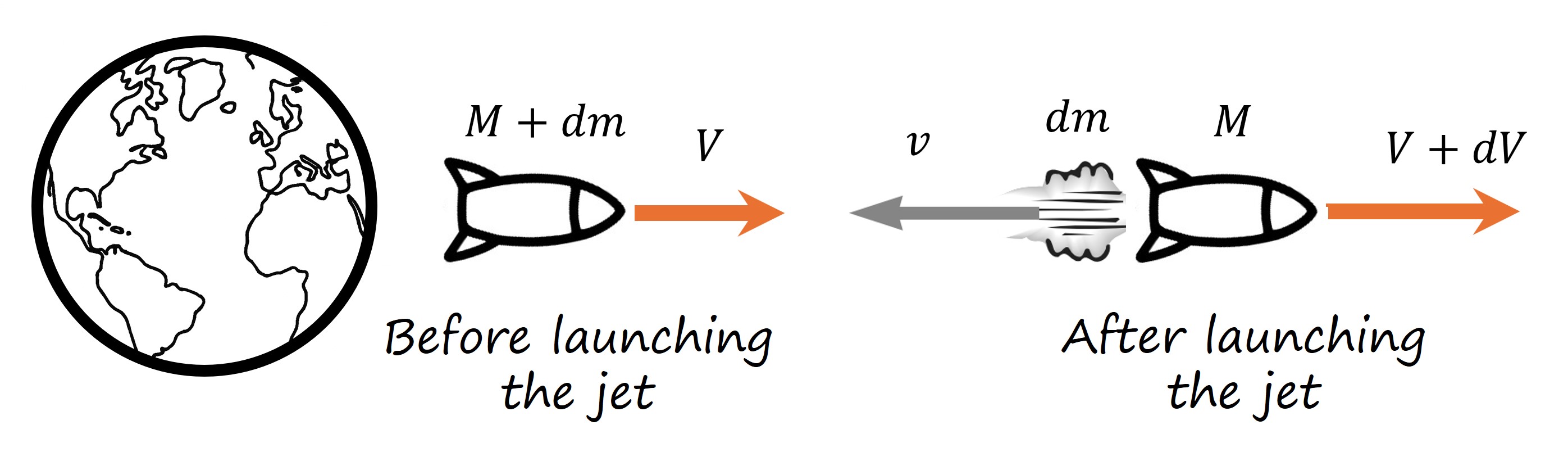}
  \caption{The propulsion of a rocket is due to the emission of fuel through the nozzle, which causes the speed to increase by an amount $\Delta V$}
\end{figure}

After a short interval, the rocket expels a jet of burned fuel through its nozzle, with mass $dm$. As a result, its velocity increases by an amount $dV$, so that the final velocity is $V + dV$. Consequently, the final momentum of the rocket is

\begin{equation}
p_f = M(V + dV) + dm\,v.
\tag{A2}
\end{equation}

By conservation of linear momentum, we must have $p_i = p_f$, so that

\begin{equation}
(M + dm)V = M(V + dV) + dm\,v.
\tag{A3}
\end{equation}

After some algebra, we obtain

\begin{equation}
M\,dV = (V - v)\,dm.
\tag{A4}
\end{equation}

Defining $v_e \equiv V - v$ as the velocity of the expelled gases relative to the rocket, or exhaust velocity, we obtain

\begin{equation}
M\,dV = v_e\,dm,
\tag{A5}
\end{equation}

where $v_e$ is constant. On the other hand, we know that the variation $dM$ in the mass of the rocket is negative and is given by $dM = M - (M + dm) = -dm$, so that $dm = -dM$. Substituting this result into Eq.~(A5), we obtain

\begin{equation}
M\,dV = -v_e\,dM \quad \Rightarrow \quad dV = -v_e \frac{dM}{M}.
\tag{A6}
\end{equation}

Integrating the left-hand side between the initial velocity $v_0$ and the final velocity $v$, and the right-hand side between the initial mass $M_0$ and the final mass $M$:

\begin{equation}
\int_{v_0}^{v} dV = -v_e \int_{M_0}^{M} \frac{dM}{M}.
\tag{A7}
\end{equation}

Finally, we obtain the so-called \textit{Tsiolkovsky rocket equation}, which is the foundation of astronautics and was originally derived in 1903 by the Soviet physicist Konstantin Tsiolkovsky:

\begin{equation}
\Delta v = v - v_0 = v_e \ln \left( \frac{M_0}{M} \right).
\tag{A8}
\end{equation}

This equation can also be written as

\begin{equation}
\frac{\Delta v}{v_e} = \ln \left( \frac{M_0}{M} \right).
\tag{A9}
\end{equation}

Figure A2 shows a graph of this equation. The graph shows that large values of $M_0/M$ produce only small variations in $\Delta v / v_e$.\\

\begin{figure}[H]
  \centering
    \includegraphics[width=0.45\textwidth]{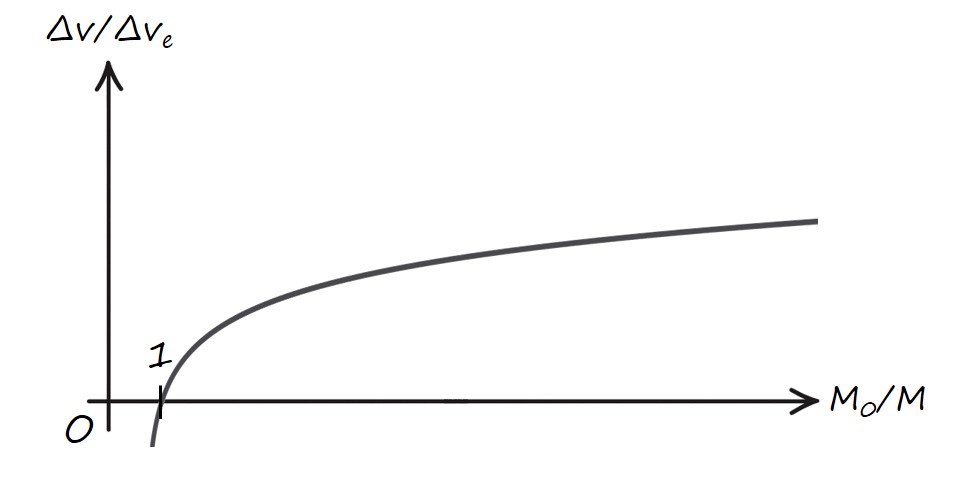}
  \caption{Graph of the basic expression for rocket propulsion. The graph shows that large values of $M_{0} / M$ produce small variations in $\Delta v/ v_{e}$.}
\end{figure}

The equation shows that the fractional change in the rocket's velocity is proportional to the natural logarithm of the ratio $M_0/M$. Since the logarithm is a slowly growing function, small increases in the rocket's velocity $\Delta v$ require enormous amounts of initial fuel. In other words, for a rocket to lift off and enter Earth orbit, almost all of its initial mass $M_0$ must be fuel. As an example, a mass ratio $M_0/M = 10^{3}$ produces only a fractional change $\Delta v / v_e \simeq 7$.

\section*{References}

[1]	Bate, Roger. R., D. Mueller D., J. White E., Fundamentals of Astrodynamics, Dover Publications Inc, New York, 1971.

\vspace{2mm}

[2]	H. Curtis D., Orbital Mechanics for Engineering Students, Elsevier, Oxford, 2005.

\vspace{2mm}

[3]	J.E. Prussing, B.A. Conway, Orbital Mechanics, 2nd ed., Oxford University Press, Oxford, 2013.

\vspace{2mm}

[4]	P.A. Tipler, Physics for Scientists and Engineers, W. H. Freeman and Company, New York, 2004.

\vspace{2mm}

[5]	H.D. Young, R.A. Freedman, Sears and Zemansky’s University physics with modern physics, 14th ed., Pearson, New Jersey, 2016.

\vspace{2mm}

[6]	C. Kittel, W.D. Knight, M.A. Ruderman, A.C. Helmholz, B.J. Moyer, Mechanics, 2nd ed., McGraw-Hill, New York, 1973.

\vspace{2mm}

[7]	R. Resnick, D. Halliday, K.S. Krane, Physics, 4th ed., John Wiley and Sons, New York, 1992.

\vspace{2mm}

[8] A. Stinner and J. Begoray, Phys. Educ. \textbf{40}, 35–45 (2005). https://doi.org/10.1088/0031-9120/40/1/002

\end{document}